\documentclass[twocolumn,aps,preprintnumbers,amsmath,amssymb,superscriptaddress,nofootinbib]{revtex4}

\usepackage{slashed}
\usepackage{epsfig,latexsym,cancel,amssymb,amsmath,verbatim,mathrsfs}
\usepackage{color}
\usepackage{graphicx,subfigure,multirow}

\usepackage{dcolumn}
\usepackage{bm}
\usepackage[all]{xy}
\usepackage{color}
\usepackage[usenames,dvipsnames]{xcolor}
\usepackage[unicode=true,pdfusetitle,
 bookmarks=true,bookmarksnumbered=false,bookmarksopen=false,citecolor=Turquoise,
 breaklinks=false,pdfborder={0 0 1},backref=false,colorlinks=true,pdfpagemode=FullScreen]
 {hyperref}
\usepackage{amssymb}
 \usepackage{longtable}
\usepackage{epsfig}
\usepackage{amsmath}
\usepackage{pstricks}
\usepackage{bbold}
\usepackage{slashed}
\usepackage{amssymb}
\usepackage{graphicx}
\usepackage{hyperref}
\usepackage{verbatim}
\usepackage{multirow}
\usepackage{chemarrow}
\usepackage{hhline}
\makeatletter

\newcommand{\beq}{\begin{equation}}
\newcommand{\eeq}{\end{equation}}
\newcommand{\bea}{\begin{eqnarray}}
\newcommand{\eea}{\end{eqnarray}}

\newcommand{\TeV}{\textrm{\,TeV}}
\newcommand{\GeV}{\textrm{\,GeV}}

\begin{document}

\preprint{
\begin{minipage}[b]{1\linewidth}
\begin{flushright}
CTPU-15-24\\
IPMU15-0209 \\
KEK-TH-1882
 \end{flushright}
\end{minipage}
}

\title{Apparent unitarity violation in high mass region of $M_{bW}$ \\ from a ``hidden" top partner at high energy colliders}

\author{Chengcheng Han}
\email[]{chengcheng.han@ipmu.jp }
\affiliation{Kavli IPMU (WPI), The University of Tokyo, Kashiwa, Chiba 277-8583, Japan}
\author{Mihoko M.\,Nojiri}
\email[]{nojiri@post.kek.jp }
\affiliation{Kavli IPMU (WPI), The University of Tokyo, Kashiwa, Chiba 277-8583, Japan}
\affiliation{KEK Theory Center, IPNS, KEK, Tsukuba, Ibaraki 305-0801, Japan}
\affiliation{The Graduate University of Advanced Studies (Sokendai),Tsukuba, Ibaraki 305-0801, Japan}
\author{Myeonghun Park}
\email[]{parc.ctpu@gmail.com}
\affiliation{Center for Theoretical Physics of the Universe, Institute for Basic Science (IBS), Daejeon, 34051, Korea}

\date{Dec.\,15th.\,2015}

\begin{abstract}
 Perturbative unitarity conditions have been playing an important role in estimating the energy scale of new physics, 
 including the Higgs mass as the most important example.
In this letter, we show that there is a possibility to see the hint of a new physics (top quark partner) indirectly by observing an ``apparent" unitarity violation in 
the distribution of invariant mass of $b$-jet and $W$-boson $(M_{bw})$ well above the mass of a top quark in a process of a heavy resonance decaying into a pair of top quarks.
\end{abstract}

\maketitle
\section{Introduction.}
Perturbative unitarity violation in a certain physics process indicates that some new physics involves in the process.
For example, Lee-Quigg-Thacker provided the upper bound of the Standard Model (SM) Higgs boson mass as $m_H< 1\TeV$ through the perturbative unitarity condition\, \cite{Lee:1977eg}.
Indeed the Higgs boson is discovered at $125\GeV$, well below that bound.
This observed Higgs mass has some issues in the SM because 
a top quark, as the most massive fundamental particle in the SM, 
causes a naturalness problem and vacuum meta-stability by its strong yukawa coupling with the Higgs boson.

Thus many new physics models have been introduced to solve the naturalness problem by controlling the Higgs mass against quantum corrections.
One natural way to remedy this issue is to introduce a ``top quark partner" (top partner) with its mass near TeV scale. 
In addition, through interactions with a top partner, a top quark also has been considered as the one of SM particles coupled to the beyond Standard Model (BSM) sector.  
Topcolour-assisted technicolour (TC2)\,\cite{Hill:1994hp},  composite Higgs scenarios\,\cite{Kaplan:1983fs} of the strong electroweak Symmetry Breaking (EWSB) models and models with warped extra dimensions\,\cite{Randall:1999ee} are classes of models predicting a heavy particle which would decay mostly into top quark and anti-top quark $(t\bar t)$.
The ATLAS and CMS collaborations have searched a heavy resonance that decays into $t\bar t$ at 8 TeV LHC\,\cite{Aad:2015fna}. 
In those searches, the LHC set a limit on the production of a massive color octet spin-1 particle (Kaluza Klein gluon) as well as a color singlet scalar. 
They excluded a color octet scalar with the mass below 1 TeV through a search of the process with four top quarks in the final state\, \cite{Aad:2015fna}. 

In this letter, we would like to point out an interesting feature in the above searches 
if a top partner is involved in the decaying process of a heavy resonance.
We argue that a high invariant mass region of a $b$-jet and a $W$-boson $(M_{bw})$ in a heavy resonance's decay process 
would be enhanced as  an ``apparent" perturbative unitarity violation effect due to a heavy top partner. 
We discuss the possibility of tracing a heavy top partner by measuring a high mass region of $M_{bw}$.

\section{ High $M_{bw}$ due to the  $W_L$ enhancement.}
\label{sec:ii}
If one describes an interaction between a new heavy particle and a pair of top quarks with a high dimension effective operator with Higgs, 
there would be a corresponding longitudinal $W$ enhancement involved in processes related to that operator through the Goldstone boson equivalence theorem.
In fact, effective operators with the Higgs field are very common in various BSMs. 
Particularly when a new particle is $SU(2)_L$ singlet, high dimensional operators with top quark pair can be written by inserting the Higgs field.

Assuming the new particles is spin-1 $(G)$, one of effective operators would be:
\begin{eqnarray}
\mathcal{L}\supset \frac{c_t}{\Lambda^2}G_\mu\, (\bar{Q}_L \tilde{H})\,\gamma^\mu \,(\tilde{H}^\dagger Q_L) \, ,
\label{eq:eff1}
\end{eqnarray}
where $H$ is the SM Higgs and $Q_L$ are the SM third generation left-hand quark. $\tilde{H}$ is $SU(2)_L$ conjugate Higgs doublet. 
If a new particle is spin-0 $(\sigma)$, one can write an effective operator:
\begin{eqnarray}
\mathcal{L}\supset \frac{c_t}{\Lambda}\sigma\, \bar{Q}_L \,\tilde{H} \,t_R\, .
\label{eq:eff2}
\end{eqnarray}
Effective operators would induce an perturbative unitary violation in a typical process 
\beq
t(p_1)+ \sigma/G(p_2) \rightarrow W(p_3)+ b(p_4)\, .
\label{eq:process}
\eeq
A perturbative unitarity violation is induced by a longitudinal $W$ contribution.
In the high momentum limit, as $\epsilon_L$ becomes proportional to $p_3^\mu/m_W$, the amplitude increases with a collision energy of  $\sqrt{\hat s}$ resulting in a perturbative unitarity violation\footnote{Unitarity violation effects including Dark matter simplified models and the bound on new physics can be found in \cite{unitarity}.}. 
Such process is promising because the corresponding cross section deviates from expected SM backgrounds strongly even though the scale of the interaction is far below the cut off scale.
In following sections, we show how above effective operators could arise in the low energy limit after integrating out a ``top partner" 
in various simplified models. We can find those simplified models in the extra dimensional model or a supersymmetric one.

\subsection{Spin-1 case}
In this subsection, we provide a simplified model by extending the SM with two new particles: one is a massive spin-1 color octet gauge boson $G$ and the other is a vector-like $SU(2)_L$ singlet top partner $T$.
 This simplified model can be embedded naturally in extra dimension models where the spin-1 color octet is the gluon Kaluza Klein mode ($KK$) and the $T$ is the $KK$-mode of a SM top quark.
The interaction Lagrangian is:
\begin{eqnarray}
 \mathcal{L} \ni && g_s G_\mu \bar{T}\gamma^\mu T-m_T \bar{T}T  \nonumber \\
 &&-( y \bar{Q}_L \tilde{H} T_R+y_t \bar{Q}_L \tilde{H} t_R + h.c.) ,
\label{eq:Gsimp}
\end{eqnarray}
with the color and $SU(2)_L$ indices suppressed for the simplicity.  
Here we adopt a model independent approach by considering $\left(y, m_T, M_G\right)$ as free parameters.
After the electroweak symmetry breaking, the top quark will be mixed with the top partner $T$ and a mass matrix can be summarized as:
\begin{eqnarray}
 \left( \begin{array}{cc} \bar{t}_L & \bar{T}_L \end{array}\right)
 \left( \begin{array}{cc}  y_t \frac{v}{\sqrt{2}} & y \frac{v}{\sqrt{2}} \\ 0 & m_T
 \end{array} \right)\ \left( \begin{array}{c} t_R \\ T_R \end{array}\right) \, .
\label{eq:neutmass}
\end{eqnarray}
This mass matrix can be diagonalized by bi-unitary transformation $U_L M U_R^\dagger=M_{\textrm{diag}}$,
where a relation between mass and weak eigenstates is
\begin{eqnarray}
 \left( \begin{array}{cc} t^\prime_{L,R} \\ T^\prime_{L,R} \end{array}\right)
 = \left( \begin{array}{cc}  \cos \theta_{L,R} &  \sin\theta_{L,R} \\ -\sin\theta_{L,R} & \cos \theta_{L,R}
 \end{array} \right)  \left( \begin{array}{cc} t_{L,R} \\ T_{L,R} \end{array}\right) \, .
\label{eq:neutmass}
\end{eqnarray}
Here $t^\prime$ and $T^\prime$ are the mass eigenstates with mixing angles of $\theta_L $ and $\theta_R$.
For $m_T\gg y_i v$, we can approximate the mixing angles,
\begin{equation}
\sin\theta_L\sim \frac{y v}{\sqrt{2}m_T},\quad\sin\theta_R\sim \frac{m_t}{m_T}\sin\theta_L\, .
\end{equation}
Constraints on mixing angles can be obtained from various precision measurements including $R_b$. 
The current limit is $\sin \theta_{L}\lesssim 0.1$\ for $m_{T'} > 1$ TeV \cite{Aguilar-Saavedra}. 
After integrating out $T^\prime$, an effective operator between SM-like top quark and a heavy resonance $G$ would be
\begin{equation}
\frac{y^2 g_s}{m_T^2}G_\mu (\bar{Q}_L \tilde{H})\gamma^\mu (\tilde{H}^\dagger Q_L) \, ,
\label{eq:effG}
\end{equation}
involving only left handed SM-like top quark as in Eq.(\ref{eq:eff1}).
The interaction between $G$ and a right handed SM-like top quark would be suppressed by the factor of $m_t/m_T$ compared to the left handed case.

\subsubsection{Gauge symmetry and unitarity restoration}

The effective operator in Eq.\,(\ref{eq:effG}) would induce a perturbative unitary violation in the high energy limit of some processes.
Here we show a process which violates a perturbative unitarity with described by this effective operator and how one can restore a unitarity by considering the contribution of a top partner.
A process that we consider here\footnote{We use $t$ and $T$ for the mass eigenstates from now on.},
\beq
t(p_1)+ G(p_2) \rightarrow W(p_3)+ b(p_4)\, ,
\label{eq:process}
\eeq 
with the effective operator in Eq.(\ref{eq:effG}). In this process, we encounter a violation of the perturbative unitarity through the $W_L$ contribution as we discussed in earlier section.
\begin{figure}[t!]
\includegraphics[width=0.45\textwidth]{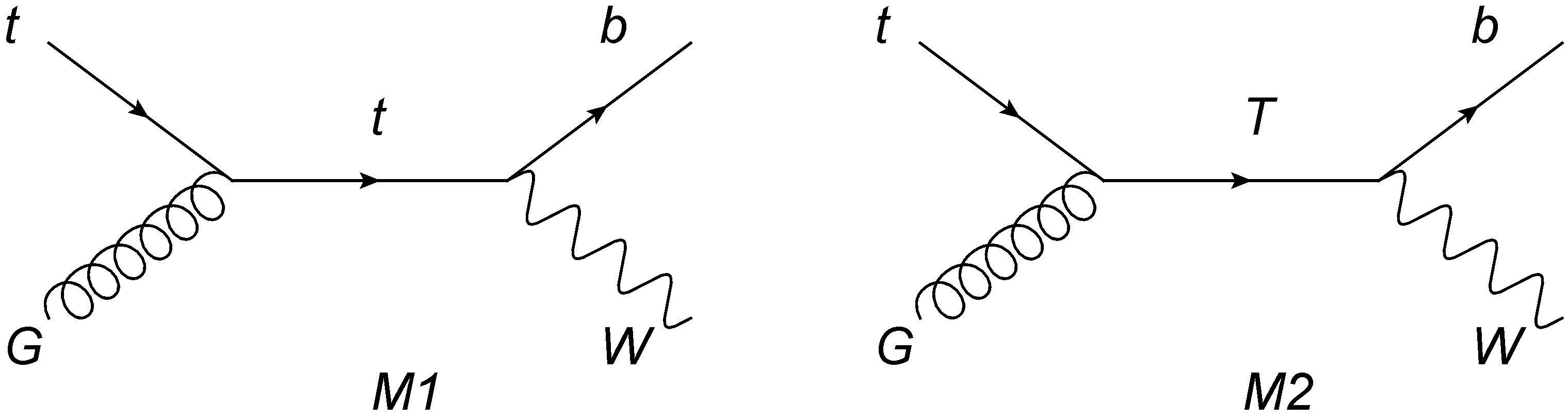}
\centering
\caption{\label{fig:fey} Feynman diagrams for the process $t+ G \rightarrow W+ b$.}
\end{figure}
But in the high energy collision limit of $\sqrt{\hat s} \gg m_T$, we should consider the original interactions of Eq.\,(\ref{eq:Gsimp}) 
with both a top quark $t$ and a top partner $T$ involved in s-channel. The corresponding feynman diagrams are shown in Fig.(\ref{fig:fey}) where the interactions among $(G,\,t,\,t)$, $(G,\,t,\,T)$ and $(T,\,W,\,b)$ are induced by mixings.
If we denote the matrix element with a top quark in s-channel as $\mathcal{M}_1$ 
and the matrix element with a top partner as $\mathcal{M}_2$:
\bea
\mathcal{M}_1 &&\propto \bar u_b\frac{\slashed{p}_3}{m_w}P_L\frac{1}{\slashed{p}_1+\slashed{p}_2-m_{t}}  \gamma_\sigma P_L u_{t} \epsilon^\sigma_G\nonumber \\
&& \propto \frac{1}{m_w} \bar u_b P_R\left(\frac{m_{t}}{\slashed{p}_1+\slashed{p}_2-m_{t}} +1\right) \gamma_\sigma P_L u_{t} \epsilon_G^\sigma\, ,\nonumber \\
\mathcal{M}_2 &&\propto - \bar u_b\frac{\slashed{p}_3}{m_w}P_L\frac{1}{\slashed{p}_1+\slashed{p}_2-m_{T}}\gamma_\sigma P_L u_{t} \epsilon_G^\sigma\, \nonumber \\
&& \propto -\frac{1}{m_w} \bar u_b P_R\left(\frac{m_{T}}{\slashed{p}_1+\slashed{p}_2-m_{T}} +1\right)\gamma_\sigma P_L u_{t} \epsilon_G^\sigma \, . \qquad
\eea
 The relative minus sign of $\mathcal{M}_2$ to $\mathcal{M}_1$ is from the mixing matrix in Eq.\,(\ref{eq:neutmass}).
 As we see in the above equation, when a collision energy $\sqrt{\hat s}$ is small compared to the mass of a top partner $m_T$, 
 $\mathcal{M}_2$ is highly suppressed by a factor of $m_t/m_T$ compared to  $\mathcal{M}_1$ resulting in 
 total matrix element becoming the case of the effective operator in Eq.\,(\ref{eq:effG}). 
The constant term in the bracket of $\mathcal{M}_1$ contributes a term proportional to $\sqrt{s}$ which seems like to invoke the perturbative unitarity violation.
But in the case of $\sqrt{\hat s} \gg m_T$, a contribution from $\mathcal{M}_2$ to the total matrix element of eq.\,(\ref{eq:process}) cancels this constant term:
\beq
\mathcal{M} \propto \bar u_b P_R\left(\frac{m_{t}}{\slashed{p}_1+\slashed{p}_2-m_{t}}-\frac{m_{T}}{\slashed{p}_1+\slashed{p}_2-m_{T}}\right)\gamma_\sigma P_L u_{t} \epsilon_G^\sigma\, .
\label{eq:full}
\eeq
The constant term in $\mathcal{M}_i$ cancels each other and thus perturbative unitarity is restored\footnote{Similarly in SM $t+ H \rightarrow W+ b$ process, the divergency in the S-channel of $t$ quark is canceled by a t-channel mediated by $W$.}.

\subsubsection{Tracing the effect from massive top partner at colliders}

\begin{figure}[t!]
\includegraphics[width=0.45\textwidth]{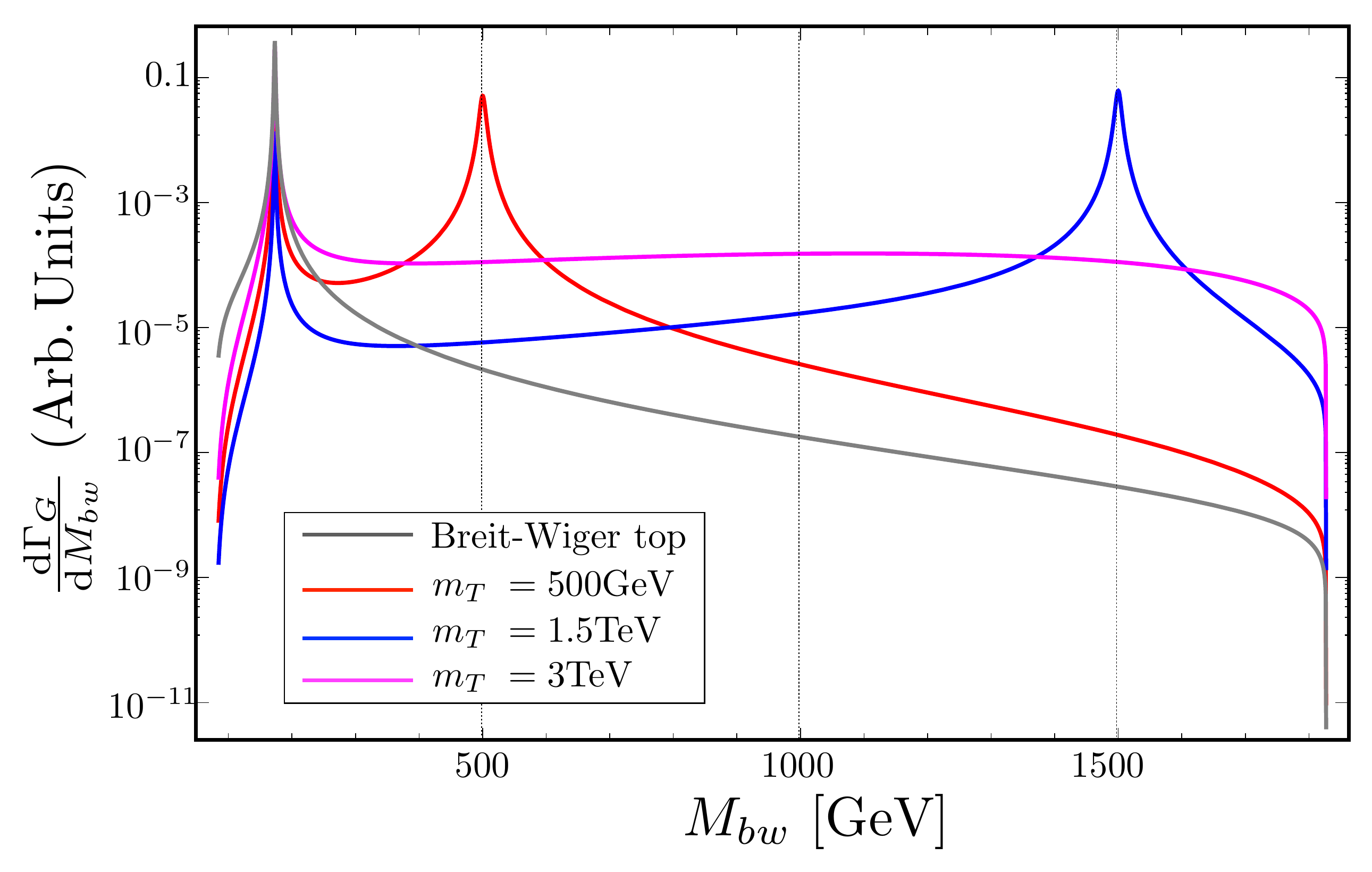}
\centering
\caption{\label{fig:offshellT} Invariant mass distributions $M_{bw}$ for various top partner masses $m_{T}$. Gray line shows the Breit-Wigner approximation of the
distribution of a top quark's mass. Here we set the mass of a color octet resonance $(m_{G})$ to be 2 TeV. }
\end{figure}

In this section, we point out that an ``apparent" unitary violation indicates the existence of a heavy top partner 
together with a discovery of a resonance $G$ if one can design collider analyses properly. 
The relevant process would be the production of $G$ and its corresponding decay.
In Fig.(\ref{fig:offshellT}) we show an invariant mass distribution of $b$-jet and $W$ boson $(M_{bw})$ from $G$ decays with various choices of $m_T$. 
When a top partner is lighter than a resonance $G$, $ M_{bw}$ distribution becomes similar to the Breit-Wigner distribution in the high mass tail of two intermediate peaks, 
one from SM-like top quark and the other from a top partner.
If a top partner is heavier than $G$, the ``peak" from a top partner mass distribution will be moved beyond the location of $M_{G}$, resulting in ``apparent" unitary violation. 
Thus the high $M_{bw}$ region can prove the existence of some ``hidden" particles, a massive top partner.

\begin{table*}[t!]
\centering
\begin{tabular}{|c||c|c|c|c|c|}
\hline    Analysis cut   & MC1 & MC2  & $\left(t\bar{t}bW\bar{b}W\right)$  & $\left(bW\bar{b}Wb\bar{b}\right)$    \\
\hline      $P_T$ requirements ($p_T^{\ell,j} > 50\GeV$)   &  0.22 & 0.19 & 0.18  &  5.5   \\
\hline    Heavy flavored jet tagging (one high $P_T$ top and four b-jet)  & 0.051  &0.050 &0.0024 &  $10^{-3}$ \\
\hline     $\textrm{M}^{(\max)}_{(b,\ell)} > 170\GeV$   & 0.012 & 0  & $10^{-4}$  & $<10^{-4}$\\
\hline
\end{tabular}
\caption{Cut flow for a signal process with and without off-shell effect(MC1,MC2) and backgrounds. Numbers in this table are cross sections after applying corresponding cuts in the unit of a femto barn (fb).
}
\label{tab:cut}
\end{table*}
Motivated by the fact that $KK$-gluon does not couple to the pair of gluon \cite{Arai:2013tsa} while it interacts mostly with top quark pairs, 
we consider $G$ pair production followed by the 
four $b$-jets and four $W$-bosons through intermediate top quarks at the LHC.
To reduce SM QCD backgrounds, we focus on a di-lepton channel with four $b$-jets tagging where leptons are expected from $W$ boson decays.
In the SM, the invariant mass of a lepton and b-jet ($m_{(b\ell)}$) from a top quark decay is less than $m_t \sqrt{1-m_W^2/m_t^2} \sim 153\GeV$.
Therefore by looking for the high mass region of $m_{(b\ell)}$, one can check the effect of a hidden top partner involved in the $G$ decaying process.
We define the following variable:
 \beq
\textrm{M}^{(\max)}_{(b,\ell)}= \max\limits_{i =1,2}\left(\min\limits_{j=1,\cdots 4}\{m_{(b_j,\ell_i)} \} \right)\, ,
 \eeq
 where we pair the $b$-jet which provides the minimum invariant mass with each lepton and we take the larger value out of two invariant mass.
To see effects from a heavy top partner ($m_T \gg m_{G})$ in the production of $G$ pair, we use the parton level Monte Carlo (MC) simulations 
of MadGraph\_aMC@NLO\, \cite{mad} and Pythia 6.4\, \cite{pythia}.
To see the effects from heavy top partner more clearly we perform MC simulations with / without effects from a heavy top partner. 
\begin{itemize}
\item[MC1:] To include effects from a top partner on $G$ decay, 
we simulate the $\left(2\to4\times3\right)$ process with each $G$ decaying into $b$-jet and particles from $W$-boson decaying.
\item[MC2:] MC simulation for the process of $\left(2\to 4, \, 4\to 12\right)$ by requiring $G$ decaying into $t\bar t$ 
followed by the successive decays of ``on-shell" top quarks\footnote{In various multi-top quarks searches at the LHC, most collider analyses are based on the MC simulations with``on-shell" top quarks to avoid the complications in generating many body final states with MC.}.
\end{itemize} 
\begin{figure}[t]
\includegraphics[width=0.42\textwidth]{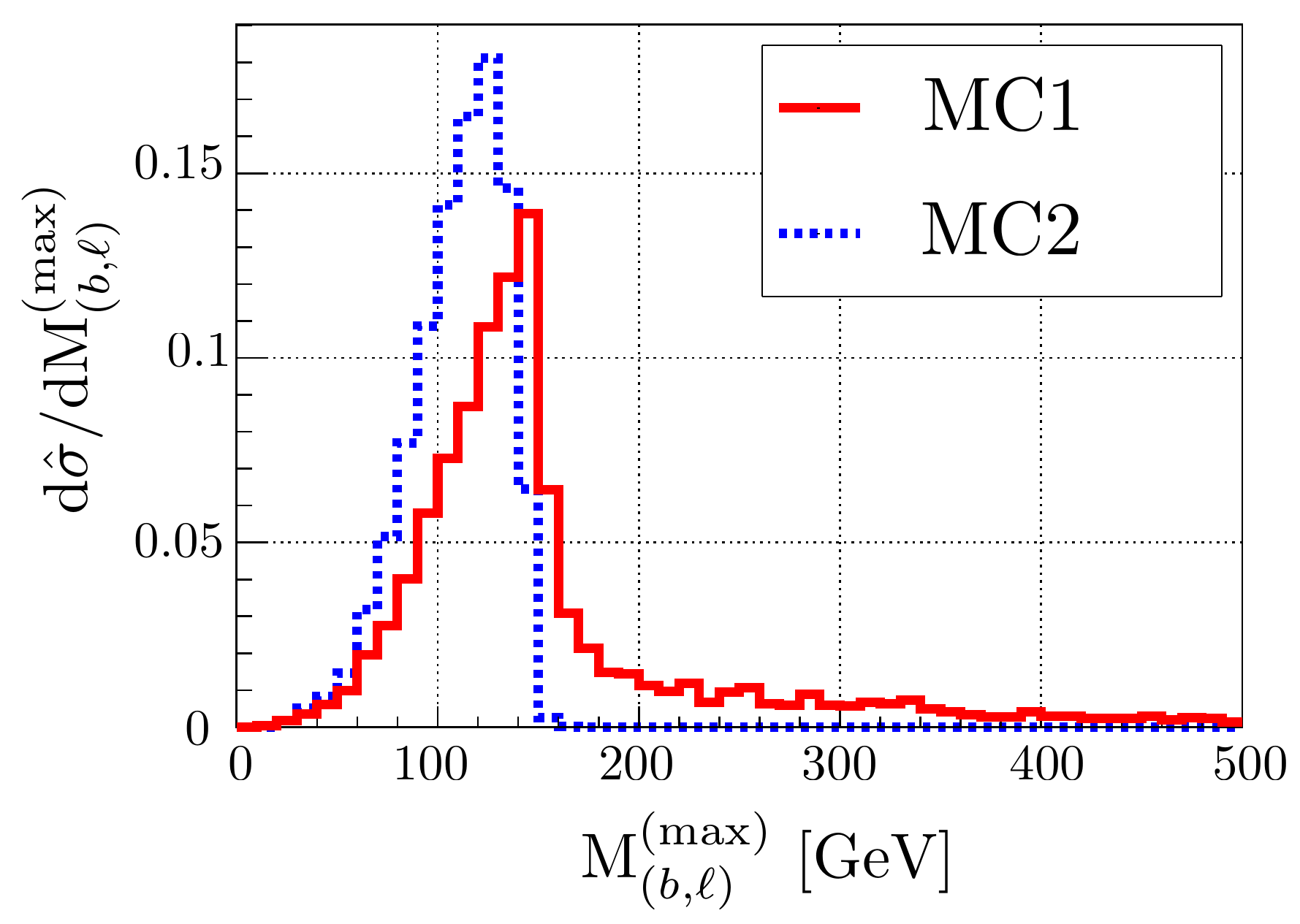}
\centering
\caption{\label{fig:dis}  Normalized distributions of $\textrm{M}^{\rm{max}}_{(b,\ell)}$ from two different Monte Carlo simulations. 
One corresponds to MC1, a full process of $G \to (b,\,\bar b,\, W^+,\, W^-)$ (red solid line). 
We also simulate MC2, the $G \to (t, \, \bar t)$ and successive decays of the top quarks to $b$-quark and $W$ boson (blue dotted line). 
Parameters in this simulation are described in the text. }
\end{figure}
By requiring ``on-shell" top quarks from $G$ decay [MC2], we can remove the effect of a heavy top partner in MC simulation. 
In fact, the result of MC2 would be similar to a case where $G$ interacts only with right handed top quarks $t_R$. 

The parameters  we use here are $M_{G}=1.8\TeV$, $m_T=3.5\TeV$, $y=2.0$ with mixing angle $\sin \theta_L=0.1$ which corresponds to a cross section around $2.5\,\textrm{fb}$ in the leading order QCD at the $14\,\TeV$ LHC.  
The $\textrm{M}^{(\max)}_{(b,\ell)}$ distribution from MC2 has a sharp end point below a mass of  top quark 
while it develops excess beyond the top mass in MC1 where
top partner involves the decay of $G$ as in FIG.\,(\ref{fig:dis}). Thus we utilize $\textrm{M}^{(\max)}_{(b,\ell)}$ as a key observable 
in tracing the existence of a top partner.

For the analysis of $14\TeV$ LHC with $3\textrm{ab}^{-1}$ luminosity, we focus on a signal region of (1 top-jet, 3 $b$-jet, 2 $\ell$) final states 
by requiring one top-jet tagging with a tagging/fake efficiency 0.5/0.01 \cite{CMS:2009lxa} and also four $b$-jet tagging with a tagging efficiency as $0.8$. 
Note that $b$-jet tagging can be performed on a top tagged jet. 
Since we consider a heavy resonance $G$, we require $p_T>50\GeV$ on visible particles (jet and leptons) and  $p_T^{ \rm{top-jet}}> 500\GeV$. 
We consider two major backgrounds $\left(t\bar{t}W\bar{b}Wb\right)$ and $\left(bW\bar{b}W b\bar{b}\right)$ with a leptonically decaying $W$ according to our selection on the signal region. 
The efficiencies after the sequential cuts are summarized in Tab.\,(\ref{tab:cut}).
According to our simulation, we will have $\mathcal{O}(30)$ events purely from the effect of very heavy top partner while we will have negligible number of events 
for on-shell top decay [MC2]. 
Although we perform a parton level analysis where we neglect smearing effects from detector responses, our result is encouraging because backgrounds are significantly reduced compared to the signals.

We note that conventional searches for a heavy resonance in a top quark pair production channel require a top-jet mass around the top quark's pole mass through various top-tagging method\, \cite{Thaler:2008ju,Kaplan:2008ie} . 
These top taggers will remove the effects from top partner and a separated analysis is required to measure effects from a top partner.

\subsection{Spin-0 case} \label{sec:sgluon}
\begin{figure}[th]
\includegraphics[width=0.47\textwidth]{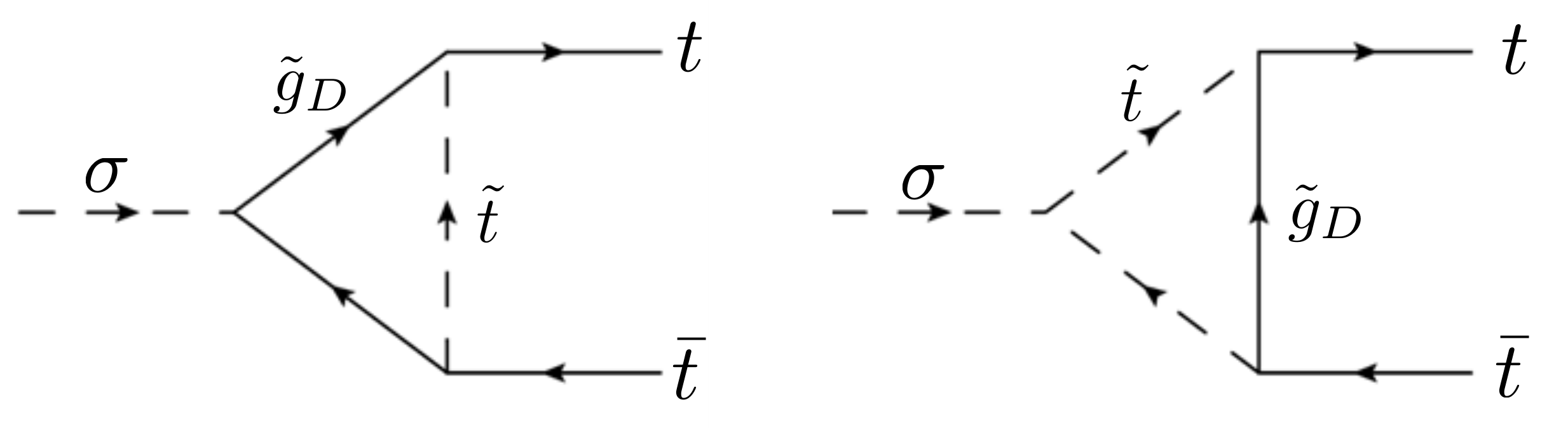}
\centering
\caption{\label{fig:diag} Effective operators in our simplified model set up between a colour octet scalar $\sigma$ and top quarks in a sec.\,(\ref{sec:sgluon}). }
\end{figure}
In this section, we present another model of color octet scalar particle. In $N =1/N =2$ hybrid Supersymmetric model\, \cite{Choi:2008ub,coupling}, a gluino could be a Dirac fermion and  a scalar gluon ($\sigma$) is predicted as the partner of the $N=2$ multiplet.
The corresponding interactions are:
\begin{eqnarray}
&& \mathcal{L}_{\tilde{g}_D \tilde{g}_D \sigma}=-\sqrt{2}i g_s f^{abc} \overline{\tilde{g}}^a_{D_L}\tilde{g}^b_{D_R} \sigma^c +h.c. \, ,\\
&&  \mathcal{L}_{\sigma \tilde{q} \tilde{q}}= -g_s M_3^D\left[\sigma^a \frac{\lambda^a_{ij}}{\sqrt{2}} \sum_q (\tilde{q}^*_{Li} \tilde{q}_{Lj}-\tilde{q}^*_{Ri} \tilde{q}_{Rj})\right]\, ,~~
\end{eqnarray}
where $\tilde{g}_D $ is a Dirac gluino with a mass $M_3^D$ and $\tilde{q}_{L,R}$ are squark gauge eigenstates.
\begin{figure}[t]
\includegraphics[width=0.45\textwidth]{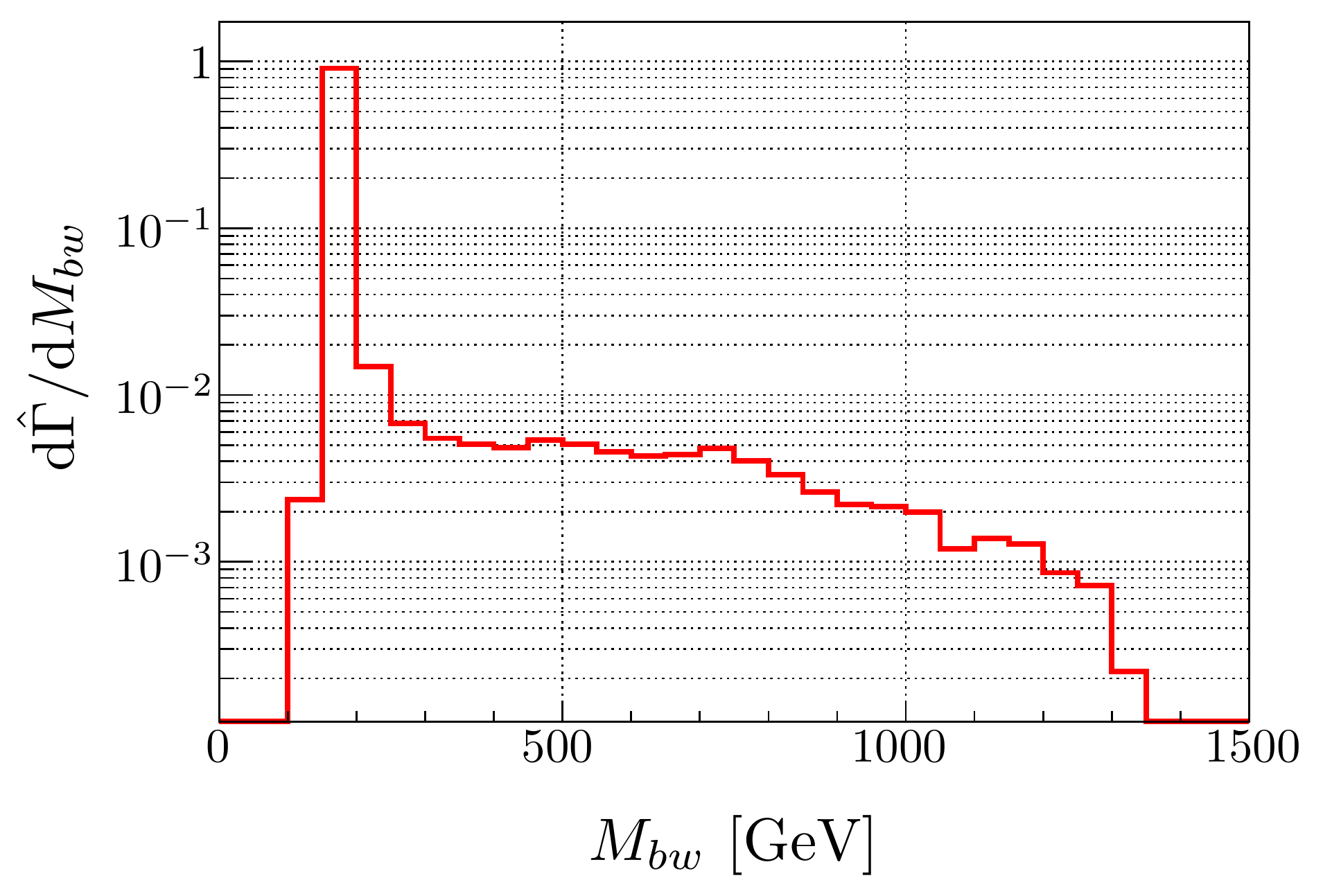}
\centering
\caption{\label{fig:sigma} {Invariant mass distribution $M_{bw}$ from $\sigma$ decay. The parameters in MC simulations are $M_\sigma=1.5\TeV$ and 
$\left(a_L,a_R\right)=\left(0.1, 0\right)$. }}
\end{figure}
Although $\sigma$ has no direct coupling with SM fermions, it can decay into $\bar{q} q$ through the squark-gluino loop as in FIG.\,(\ref{fig:diag}). 
When a Dirac gluino and squarks are heavy compared to $\sigma$, we can write down a effective operator between $\sigma$ and top quarks with generalized couplings by integrating out those heavy particles \cite{coupling}:
\begin{eqnarray}
\mathcal{L}_{int}=\sigma\, \bar{q}\left[a_L P_L+ a_R P_R\right]q + h.c.\, .
\label{eq:simp0}
\end{eqnarray}
The $a_L$ and $a_R$ are induced by right and left squark with gluino loop respectively. 
Thus depending on the mass term of left and right squark as well as the mixing term, the coupling $a_{L/R}$ can have non-zero value.
As pointed out in \cite{Choi:2008ub}, $\sigma$ decays mostly into the pair of a top quark since an effective interaction between $\sigma$ and quark pair is proportional to the mass of a quark, coming from the quark $L/R$ mixing and squark $L/R$ mixing term with the Higgs field in a superpotential and soft SUSY breaking terms. 
Thus one can write the effective operator in Eq.\,(\ref{eq:simp0}) as
\begin{eqnarray}
\mathcal{L}_{int} \ni c_R \frac{y_t}{\Lambda_1}\sigma\, \bar{Q}_L \tilde{H}t_R + c_L\frac{y_t}{\Lambda_2} \sigma\, \bar t_R \tilde{H}^+ Q_L + h.c.\, ,
\end{eqnarray}
where the $\Lambda_i$ is the mass scale of the stops or gluino, $y_t$ is a a top yukawa coupling and $c_{L/ R}$ is a coupling constant which is proportional to $a_{L/R}$.
This isospin violating effective operator will cause the enhancement in the high $M_{bw}$ region from an enhancement in a longitudinal polarization of $W_L$ as in Sec.(\ref{sec:ii}).
We show a numerical result of this phenomena in Fig.\,(\ref{fig:sigma}). 

With $m_{\tilde q}, m_{\tilde g} \gg m_\sigma$, $\sigma$ decays into a pair of gluon and a pair of top quark. 
Since we are interested in the decay mode of $\sigma \to t \bar t$, the competing decay mode $\sigma \to g g$ should be suppressed which occurs in the parameter space of $m_{\tilde q} \ge |M_3^D|$.
In this region, a production cross section for a single sgluon  $\left(pp\to \sigma\right)$ is suppressed compared to a pair production $\left(pp \to \sigma \sigma^*\right)$ more than $\mathcal{O}(10) - \mathcal{O}(100)$ depending on $m_\sigma$ \cite{Choi:2008ub}.
Thus its LHC (hadron collider) phenomenology would be similar to the case of spin-1 resonance $\left(G\right)$ in the previous section. 
For a 1.5 TeV sgluon, the leading order cross section is around 4fb and 
the number of events in the high mass region is expected to be around $\mathcal{O}(20)$ if we apply same cuts since efficiences of analysis cuts in Tab.\,(\ref{tab:cut}) are not sensitive to the spin of a heavy resonance.
\\
\section{Conclusion}

We consider a case where a heavy resonance couples to a pair of top quarks in the isospin violating simplified models, 
which can be realized in various interesting BSM scenarios including an extra dimensional or a supersymmetric model. 
We first show how these simplified models invoke a perturbative unitarity violation by identifying the corresponding effective operators for the interaction between a heavy resonance and top quarks.
We find that the Higgs field in these effective operators provides an enhancement of $W_L$ in the high energy limit through the unitarity violation
in the low energy effective theory.
After we notice that a top partner introduces the Higgs in those problematic effective operators, 
we examine the possibility of tracing a top partner by measuring the high invariant mass region of $b$-jet and $W$-boson $(M_{bw})$. 
Through Monte Carlo simulations, we confirm that we can observe the effect from a very heavy top partner with a high luminosity LHC (HL-LHC) of $3\textrm{ab}^{-1}$
even if the LHC can not directly produce it.
Finally  we need non-conventional analyses for a heavy resonance search in multi-top channels, since top-jet tagging in those analyses rejects events which are crucial in identifying the existence of a top partner.

\acknowledgements
CCH is thankful to Michihisa Takeuchi, Masahiro Ibe and Shigeki Matsumoto for useful discussions.
MP appreciates comments from Kiwoon Choi, Seong Youl Choi and P.\,Ko.
MP is supported by IBS under the project code, IBS-R018-D1.
CCH and MMN are supported by World Premier International Research Center Initiative (WPI Initiative), MEXT, Japan.
MMN is also supported by Grant-in-Aid for Scientific research Nos.\,23104006 and 26287039.

\end{document}